\documentclass[a4paper]{spie}  

\usepackage[dvipsnames]{xcolor}
\usepackage{amsmath,amsfonts,amssymb}
\usepackage{graphicx}
\usepackage[colorlinks=true, allcolors=blue]{hyperref}
\usepackage{tensor}
\usepackage{tcolorbox}
\usepackage{amsbsy}
\usepackage{textcomp}
\usepackage{xfrac}
\usepackage{booktabs}
\usepackage{pifont}
\usepackage{aas_macros}

\usepackage[font={small}, skip=\abovecaptionskip]{caption}

\newcommand{\vect}[1]{\boldsymbol{#1}}
\newcommand{\maximize}[1]{\underset{#1}{\operatorname{maximize}}~~}
\newcommand{\subjectto}{\operatorname{subject~to}~~}
\newcommand{\fraunhofer}[2]{\mathcal{P}_{#1}\{#2\}}
\newcommand{\coro}[2]{\mathcal{P}^{-1}_{#1}\{M(\vect{k}) \cdot \mathcal{P}_{#1}\{#2\}\}}
\newcommand{\Real}[1]{\Re\{#1\}}
\newcommand{\Imag}[1]{\Im\{#1\}}

\newcommand{\cmark}{{\color{ForestGreen}\ding{51}}}
\newcommand{\xmark}{{\color{red}\ding{55}}}

\title{Exploiting symmetries and progressive refinement for apodized pupil Lyot coronagraph design}

\author[a]{Emiel H. Por}
\author[b]{Rémi Soummer}
\author[b]{James Noss}
\author[b]{Kathryn St.Laurent}
\affil[a]{Leiden Observatory, Leiden University, P.O. Box 9513, 2300 RA Leiden, The Netherlands}
\affil[b]{Space Telescope Science Institute, 3700 San Martin Drive, Baltimore, MD 21218, USA}

\authorinfo{Correspondence: \tt{por at strw.leidenuniv.nl}}

\begin{document}
\maketitle

\begin{abstract}
Modern coronagraph design relies on advanced, large-scale optimization processes that require an ever increasing amount of computational resources. In this paper, we restrict ourselves to the design of Apodized Pupil Lyot Coronagraphs (APLCs). To produce APLC designs for future giant space telescopes, we require a fine sampling for the apodizer to resolve all small features, such as segment gaps, in the telescope pupil. Additionally, we require the coronagraph to operate in broadband light and be insensitive to small misalignments of the Lyot stop. For future designs we want to include passive suppression of low-order aberrations and finite stellar diameters. The memory requirements for such an optimization would exceed multiple terabytes for the problem matrix alone.

We therefore want to reduce the number of variables and constraints to minimize the size of the problem matrix. We show how symmetries in the pupil and Lyot stop are expressed in the complete optimization problem, and allow removal of both variables and constraints. Each mirror symmetry reduces the problem size by a factor of four. Secondly, we introduce progressive refinement, which uses low-resolution optimizations as a prior for higher resolutions. This lets us remove the majority of variables from the high-resolution optimization. Together these two improvements require up to 256x less computer memory, with a corresponding speed increase. This allows for greater exploration of the phase space of the focal-plane mask and Lyot-stop geometry, and easier simulation of sensitivity to Lyot-stop misalignments. Moreover, apodizers can now be optimized at their native manufactured resolution.
\end{abstract}



\section{Introduction}
\label{sec:introduction}

High contrast imaging aims to directly observe exoplanets and protoplanetary disks around their bright host stars. Through indirect methods, such as radial velocity and transits, we know that most stars harbor a companion in the habitable zone\cite{borucki2011characteristics}. Spectral characterization of these planets require a planet that frequently transits its bright host star. Direct imaging on the other hand spatially separates the planet light from the star light and enables detection and characterization. This allows for the detection of for example variability induced by the rotational modulation of cloud and weather systems, and the glints off surface water with their polarization signal.

\begin{figure}[t]
\centering
\includegraphics[width=0.85\textwidth]{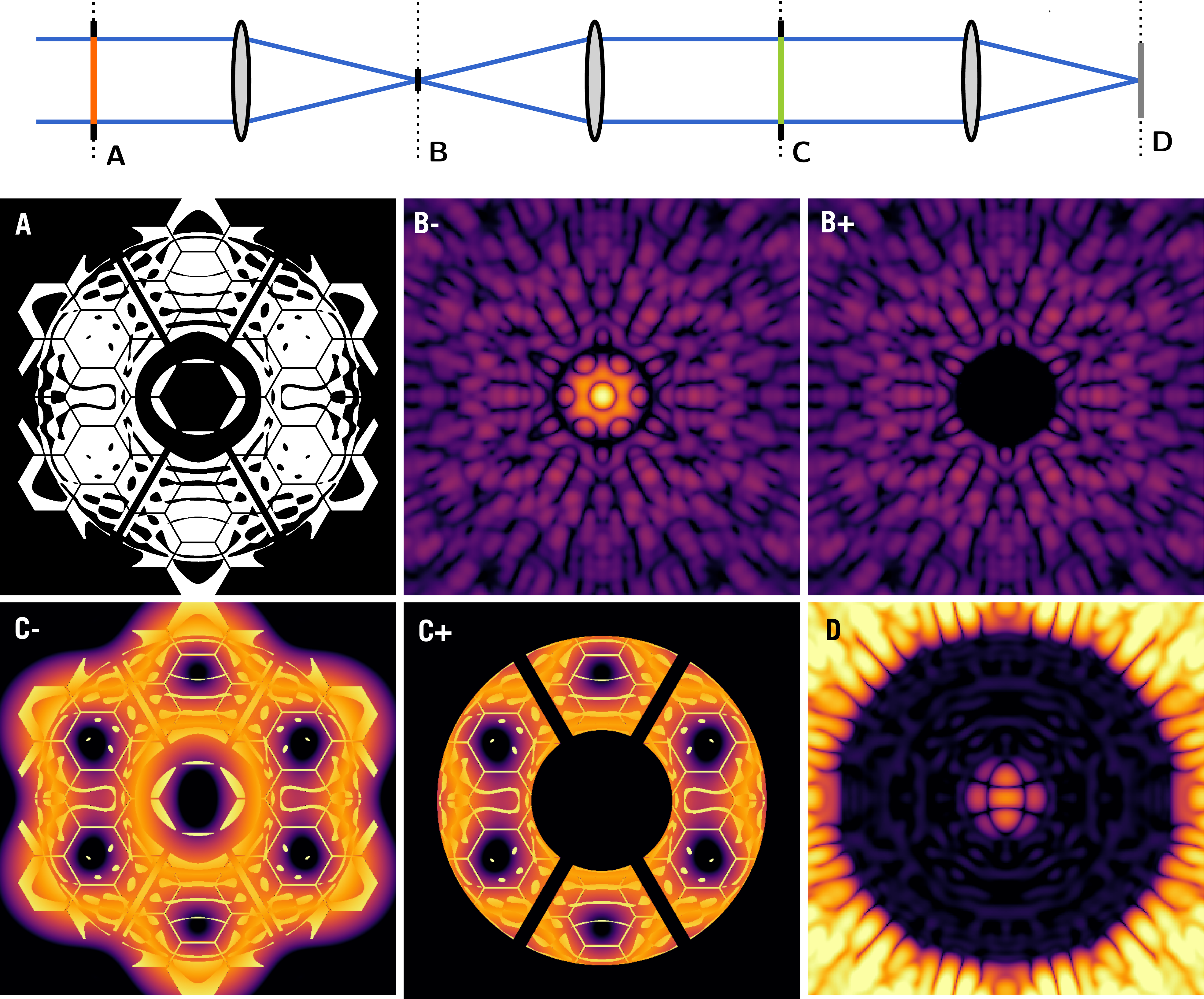}
\vspace*{3mm}
\caption{The schematic layout of the APLC with an example for light propagation through the coronagraph. The label B- means just before plane B, and B+ means just after plane B. All images are on logarithmic scales, with the B images from $10^{-5}$ to $1$, planes C from $10^{-3}$ to $1$ and plane D from $10^{-9}$ to $10^{-4}$.}
\label{fig:schematic_layout}
\end{figure}

With ground-based extreme adaptive optics systems, such as the instruments VLT/SPHERE\cite{beuzit2019sphere}, Gemini/GPI\cite{macintosh2014first} and Subaru/SCExAO \cite{jovanovic2015subaru} the direct imaging of exoplanets has become a reality. Future dedicated space-based instrumentation, such as RST/CGI\cite{spergel2013wfirst} and LUVOIR\cite{pueyo2017luvoir} extend this capability to deeper contrasts between host star and the orbiting planet. These systems use advanced coronagraphs to suppress the overwhelming star light and reveal the closeby weak planet light. A well known coronagraph is the Apodized Pupil Lyot Coronagraph (APLC) \cite{soummer2004apodized, zimmerman2016shaped}. This coronagraph combines an intricately designed apodizer mask in the pupil plane of the telescope with a focal-plane mask and Lyot stop. Figure~\ref{fig:schematic_layout} shows the optical layout of the APLC and an example of light propagating through its various masks.

This paper concerns the optimization of the masks. These masks must maximize planet throughput, while still suppress the star light. This results in large-scale numerical optimization problems that challenge current computational hardware. In particular, the large number of variables in these masks (typically $\sim1$ million variables) and constraints (typically $>50,000$ or more constraints) require a large amount of computer memory to store and operate the problem matrix, the linearized model of the coronagraphic system. Often, optimizations are carried out on large-memory servers due to these computational requirements. This paper introduces two main ways to reduce the optimization problem to more manageable levels and allow masks to be optimized at much higher resolutions than before.

Section~\ref{sec:the_full_optimization_problem} builds the full optimization problem of an APLC apodizer step by step. Section~\ref{sec:removing_symmetries} introduces the removal of point symmetry and mirror symmetries from this mathematical problem. Section~\ref{sec:progressive_refinement} introduces a new way to progressively upscale low-resolution masks. Finally, Section~\ref{sec:case_studies} shows two examples of optimized masks at high resolution. We conclude with Section~\ref{sec:conclusions}.

\section{The full optimization problem}
\label{sec:the_full_optimization_problem}

In this section, we will derive the general and complete optimization problem for a Lyot coronagraph with a complex apodizer. Writing this optimization problem out explicitly makes it easier to identify the appropriate symmetries later on.

The schematic layout of the apodized Lyot coronagraph is shown in Figure~\ref{fig:schematic_layout}. We want to find the apodizer in plane A, $\Phi(\vect{x})$, that provides the maximum throughput for an off-axis source, while simultaneously providing at least a certain raw contrast for on-axis sources. We will use the Fraunhofer operator $\fraunhofer{\lambda}{\cdot}$ as the operator for propagation of an electric field from a pupil to a focal plane through a lens. With this operator we can compute the coronagraphic image $\Psi^\mathrm{coro}_{\lambda,L}$ and non-coronagraphic image $\Psi^\mathrm{psf}_{\lambda,L}$ at wavelength $\lambda$ and using Lyot stop $L$:
\begin{subequations}
\begin{align}
\Psi^\mathrm{coro}_{\lambda,L} &= \fraunhofer{\lambda}{\coro{\lambda}{\Phi(\vect{x}) \cdot \Pi(\vect{x})} \cdot L(\vect{x})}, \\
\Psi^\mathrm{psf}_{\lambda,L} &= \fraunhofer{\lambda}{\Phi(\vect{x}) \cdot \Pi(\vect{x}) \cdot L(\vect{x})},
\end{align}
\end{subequations}
where $\lambda_0$ is the center wavelength of the light, $M(\vect{k})$ is the focal-plane mask of the Lyot stage, and $L(\vect{x})$ is the Lyot-stop mask.

For our optimization problem we will maximize the peak of the non-coronagraphic image, while simultaneously constraining the intensity of the coronagraphic image in the dark zone $D$ to be smaller than the desired raw contrast limit $10^{-c(\vect{k})}$ relative to the peak intensity of the non-coronagraphic image. This whole optimization problem reads as
\begin{subequations}
\begin{align}
\label{eq:objective_function}
\maximize{\Phi(\vect{x})} & ||\Psi^\mathrm{psf}_{\lambda_0,L}(0)||^2 \\
\label{eq:focal_plane_constraints}
\subjectto & ||\Psi^\mathrm{coro}_{\lambda_0,L}(\vect{k})||^2 \leq  10^{-c(\vect{k})} \cdot ||\Psi^\mathrm{psf}_{\lambda_0,L}(0)||^2 ~~&\forall~\vect{k} \in D \\
\label{eq:pupil_plane_constraints}
& ||\Phi(\vect{x})||^2 \leq 1 ~~&\forall~\vect{x}.
\end{align}
\end{subequations}
To clarify the nomenclature, we will refer to Eq.~\ref{eq:objective_function} and analogous equations as the objective function, Eq.~\ref{eq:focal_plane_constraints} and analogous equations as the focal-plane constraints, and Eq.~\ref{eq:pupil_plane_constraints} and analogous equations as the pupil-plane constraints.

This optimization problem is still incomplete and moreover exhibits several problems. In the next few subsections we will incrementally add more complexities and remove these remaining problems.

\subsection{Extensions}

\subsubsection{Broadband light}

Currently the contrast is only constrained at the center wavelength of our observing band. While we could replace our focal-plane constraints by similar constraints on our broadband coronagraphic image, we prefer to add constraints for each wavelength in our wavelength band independently. This ensures that we achieve the required contrast independent of the spectrum of the star. The extended optimization problem now reads
\begin{subequations}
\begin{align}
\maximize{\Phi(\vect{x})} & ||\Psi^\mathrm{psf}_{\lambda_0,L}(0)||^2 \\
\subjectto & ||\Psi^\mathrm{coro}_{\lambda,L}(\vect{k})||^2 \leq  10^{-c(\vect{k})} \cdot ||\Psi^\mathrm{psf}_{\lambda,L}(0)||^2 ~~&\forall~\vect{k} \in D, \\&&\nonumber\forall~\lambda \in [\lambda_\mathrm{min}, \lambda_\mathrm{max}] \\
& ||\Phi(\vect{x})||^2 \leq 1 ~~&\forall~\vect{x},
\end{align}
\end{subequations}
where $\lambda_\mathrm{min}$ and $\lambda_\mathrm{max}$ are the minimum and maximum wavelength in our wavelength band. Note that the monochromatic non-coronagraphic image is still used for the objective function.

\subsubsection{Lyot stop robustness}

We can extend the optimization problem to add Lyot robustness in a similar fashion. We now constrain the contrast for a set of Lyot stops. This set of Lyot stops may contain many slightly shifted versions of the nominal Lyot stop. In this way the APLC becomes robust against transverse translation of the Lyot-stop mask. Additionally we can imagine adding the nominal Lyot stop at slightly different scales to become robust against small magnification errors between the apodizer and Lyot-stop planes.

We will denote the nominal Lyot stop as $L_0(\vect{x})$ and the set of Lyot stops used for the focal-plane constraints as $\{L_1, L_2, \ldots L_N\}$. The extended optimization problem now reads
\begin{subequations}
\begin{align}
\maximize{\Phi(\vect{x})} & ||\Psi^\mathrm{psf}_{\lambda_0,L_0}(0)||^2 \\
\subjectto & ||\Psi^\mathrm{coro}_{\lambda,L}(\vect{k})||^2 \leq  10^{-c(\vect{k})} \cdot ||\Psi^\mathrm{psf}_{\lambda,L}(0)||^2 ~~&\forall~\vect{k} \in D, \\&&\nonumber\forall~\lambda \in [\lambda_\mathrm{min}, \lambda_\mathrm{max}], \\&&\nonumber\forall~L \in \{L_1, L_2, \ldots, L_N\} \\
& ||\Phi(\vect{x})||^2 \leq 1 ~~&\forall~\vect{x}.
\end{align}
\end{subequations}

\subsubsection{Low-order robustness}

Finally, we can add robustness against low-order aberrations in a similar fashion as well. For a set of modes $\{a_1(\vect{x}), a_2(\vect{x}), \ldots, a_M(\vect{x})\}$ with each a maximum amplitude corresponding to the maximum strength that we want to be robust to. An aberrated coronagraphic image can be computed with
\begin{equation}
\Psi^\mathrm{coro, aber}_{\lambda,L,\vect{\alpha}} = \fraunhofer{\lambda}{\coro{\lambda}{\Phi(\vect{x}) \cdot \Pi(\vect{x}) \cdot \sum_i \alpha_i a_i(\vect{x})} \cdot L(\vect{x})},
\end{equation}
This yields for the extended optimization problem
\begin{subequations}
\begin{align}
\maximize{\Phi(\vect{x})} & ||\Psi^\mathrm{psf}_{\lambda_0,L_0}(0)||^2 \\
\subjectto & ||\Psi^\mathrm{coro, aber}_{\lambda,L,\vect{\alpha}}(\vect{k})||^2 \leq  10^{-c(\vect{k})} \cdot ||\Psi^\mathrm{psf}_{\lambda,L}(0)||^2 ~~&\forall~\vect{k} \in D, \\&&\nonumber\forall~\lambda \in [\lambda_\mathrm{min}, \lambda_\mathrm{max}], \\&&\nonumber\forall~L \in \{L_1, L_2, \ldots, L_N\}, \\&&\nonumber\forall~\vect{\alpha} \in [-1, 1]^M \\
& ||\Phi(\vect{x})||^2 \leq 1 ~~&\forall~\vect{x}.
\end{align}
\end{subequations}
Again, we do not include the aberrated PSF in our objective function.

\subsection{Convexification and speed up}

As is, the current optimization problem is non-convex. This makes it difficult to find its global optimum. In this subsection we convexify the problem and make two other modifications that significantly speed up the optimization process.

\subsubsection{Removal of the phase piston symmetry}

The source of the non-convexity can be easily removed similar to other coronagraphs\cite{carlotti2011optimal, por2017optimal, por2020paplc} and previous methods published for the APLC\cite{zimmerman2016shaped}. The non-convexity in our optimization problem stems from the invariance under the phase piston symmetry transformation $S : \Phi(\vect{x}) \to \Phi'(\vect{x})=\Phi(\vect{x}) \exp{i\beta}$, where $\beta \in \mathbb{R}$ is an arbitrary constant. Therefore, if we have found a solution $\hat\Phi(\vect{x})$ for the apodizer, then $S\hat\Phi(\vect{x})=\hat\Phi(\vect{x})\exp{i\beta}$ is necessarily also a solution of the optimization problem. This means that the solution to the problem is non-unique and the problem therefore non-convex. We can remove this symmetry by maximizing the real part of the electric field at the peak of the non-coronagraphic image, rather than its absolute value. The choice of maximizing the real part, instead of any other linear combination of real and imaginary part is arbitrary. The optimization problem now reads:
\begin{subequations}
\begin{align}
\maximize{\Phi(\vect{x})} & \Real{\Psi^\mathrm{psf}_{\lambda_0,L_0}(0)} \\
\label{eq:focal_plane_constraints2}
\subjectto & ||\Psi^\mathrm{coro, aber}_{\lambda,L,\vect{\alpha}}(\vect{k})||^2 \leq  10^{-c(\vect{k})} \cdot ||\Psi^\mathrm{psf}_{\lambda,L}(0)||^2 ~~&\forall~\vect{k} \in D, \\&&\nonumber\forall~\lambda \in [\lambda_\mathrm{min}, \lambda_\mathrm{max}], \\&&\nonumber\forall~L \in \{L_1, L_2, \ldots, L_N\}, \\&&\nonumber\forall~\vect{\alpha} \in [-1, 1]^M \\
& ||\Phi(\vect{x})||^2 \leq 1 ~~&\forall~\vect{x}.
\end{align}
\end{subequations}
This optimization problem is now completely convex and therefore much easier to optimize. Any local optimum that we now find is the global optimum of this problem.

\subsubsection{Simplification of focal-plane constraints}

There are a few things that we do to speed up the optimization even further. The first is to simplify the focal-plane constraints. Rather than having the peak of the coronagraphic PSF appear on the right-hand side (RHS) of Eq.~\ref{eq:focal_plane_constraints2}, we change the RHS to $10^{-c(\vect{k})} I^\mathrm{peak}_{\lambda, L, \vect{\alpha}, i}$, where $I^\mathrm{peak}_{\lambda, L, \vect{\alpha}, i}$ is the peak of the non-coronagraphic PSF evaluated for a certain fixed apodizer $\Phi_i(\vect{x})$, at a wavelength $\lambda$, with Lyot-stop $L(\vect{x})$ and with aberration coefficients $\vect{\alpha}$. The optimization problem now reads
\begin{subequations}
\begin{align}
\maximize{\Phi(\vect{x})} & \Real{\Psi^\mathrm{psf}_{\lambda_0,L_0}(0)} \\
\label{eq:focal_plane_constraints2}
\subjectto & ||\Psi^\mathrm{coro, aber}_{\lambda,L,\vect{\alpha}}(\vect{k})||^2 \leq  10^{-c(\vect{k})} \cdot I^\mathrm{peak}_{\lambda, L, \vect{\alpha}, i} ~~&\forall~\vect{k} \in D, \\&&\nonumber\forall~\lambda \in [\lambda_\mathrm{min}, \lambda_\mathrm{max}], \\&&\nonumber\forall~L \in \{L_1, L_2, \ldots, L_N\}, \\&&\nonumber\forall~\vect{\alpha} \in [-1, 1]^M \\
& ||\Phi(\vect{x})||^2 \leq 1 ~~&\forall~\vect{x}.
\end{align}
\end{subequations}
The converged apodizer solution
\begin{equation}
\hat{\Phi}(\vect{x}) = \lim_{i\to\infty} \Phi_i(\vect{x}),
\end{equation}
is now found by starting with $\Phi_0(\vect{x}) = 1$, and letting $\Phi_{i+1}(\vect{x})$ be the solution of the above optimization problem. After the first iteration, the apodizer solution still violates the contrast by a factor of two for realistic scenarios for the APLC geometry. By the second iteration, our solution has practically converged. Usually we only perform two iterations for optimizing an apodizer mask.

This counter-intuitive change yields an order of magnitude improvement in the computation time. While the exact reason for this speed-up is unknown, it is likely to be caused by the internals of the used numerical optimizer.

\subsubsection{Linearized focal-plane constraints}

\begin{figure}
\centering
\includegraphics[width=0.5\textwidth]{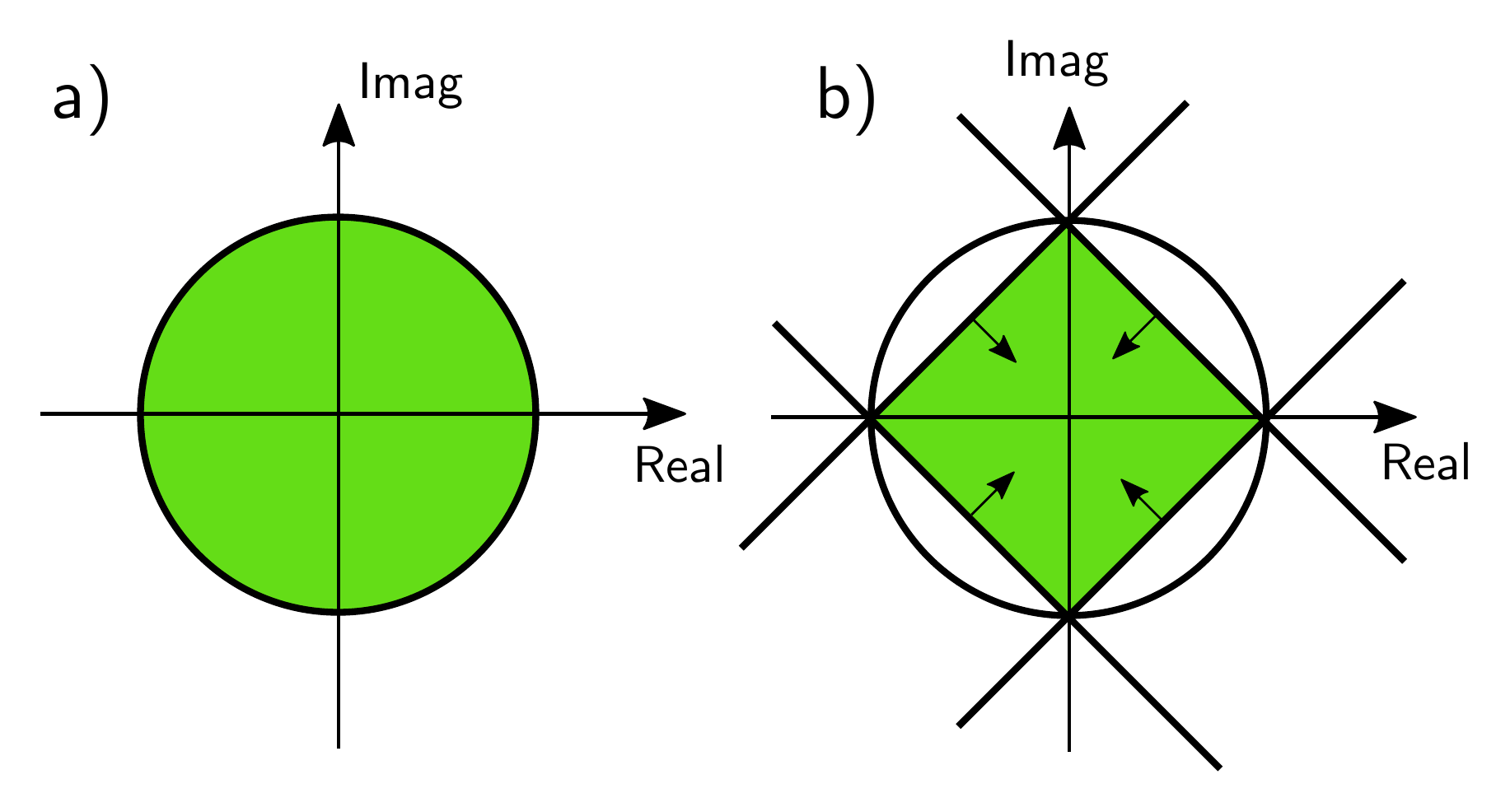}
\caption{The focal plane constraints in real and imaginary space \emph{(a)} before linearization, and \emph{(b)} after linearization. Typically the electric field is ``pushed'' towards the positive real part of this space, so having a point here ensures that no symmetry breaking occurs due to small numerical perturbations.}
\label{fig:focal_plane_constraints_linearization}
\end{figure}

By linearizing the focal-plane constraints, we can speed up the optimization even further. Linearization yields sub-optimal results but the speed improvement is often worth the extremely minor throughput loss (typically $\ll 0.1\%$). There are several ways of linearization\cite{por2017optimal}. Analogous to the optimization methods for apodizing phase plate coronagraphs \cite{por2017optimal} we choose a rotated inscribed square inside the original circular constraint. The reason for this choice is as follows. In practice we see that the electric field is ``pushed'' towards the corners of this square. In case the sides of the square lie perpendicular to real and imaginary axes, we produce an effect similar to spontaneous symmetry breaking, where a solution that would otherwise be real, gets pushed towards either positive or negative imaginary values. Therefore, we choose to have the corners of the square lie along the real and imaginary axes. This situation is shown schematically in Figure~\ref{fig:focal_plane_constraints_linearization}.

The optimization problem now reads
\begin{subequations}
\begin{align}
\maximize{\Phi(\vect{x})} & \Real{\Psi^\mathrm{psf}_{\lambda_0,L_0}(0)} \\
\label{eq:focal_plane_constraints2}
\subjectto & \pm \Real{\Psi^\mathrm{coro, aber}_{\lambda,L,\vect{\alpha}}(\vect{k})} \pm \Imag{\Psi^\mathrm{coro, aber}_{\lambda,L,\vect{\alpha}}(\vect{k})} \leq  \sqrt{10^{-c(\vect{k})} \cdot I^\mathrm{peak}_{\lambda, L, \vect{\alpha}, i}} ~~&\forall~\vect{k} \in D, \\&&\nonumber\forall~\lambda \in [\lambda_\mathrm{min}, \lambda_\mathrm{max}], \\&&\nonumber\forall~L \in \{L_1, L_2, \ldots, L_N\}, \\&&\nonumber\forall~\vect{\alpha} \in [-1, 1]^M \\
& ||\Phi(\vect{x})||^2 \leq 1 ~~&\forall~\vect{x},
\end{align}
\end{subequations}
where the $\pm$ signs have to be taken over all combinations yielding four constraints for the four sides of the square. Typically we see an order of magnitude reduction in the computation time during the optimization phase. Again, the reason for this improvement is likely due to the internals of the numerical optimizer.

\section{Symmetry reduction}
\label{sec:removing_symmetries}

Removing symmetries in our optimization problem is a powerful way of reducing computation time and memory usage. In general, symmetric optimization problems do not need to have a solution that is symmetric. Rather, points in the parameter space that are invariant under the symmetry transformation (ie. $S \vect{x} = \vect{x}$ for some symmetry transformation $S$), are critical points: either local optima or saddle points. This principle is now known as the Purkiss principle\cite{waterhouse1983symmetric}. For convex optimization problems however, the global solution is guaranteed to be symmetric. After all, if we have found the unique solution $\hat{x}$ of the convex optimization problem, its symmetric solution $\hat{\vect{x}}' = S\hat{\vect{x}}$ should also be solution. Since a convex optimization problem has a unique global optimum, the found solution must be invariant under the symmetry transformation: $\hat{\vect{x}}' = S\hat{\vect{x}} = \hat{\vect{x}}$.

Therefore, if our optimization problem exhibits a certain symmetry, we are able to remove variables and constraints, as we know the solution has to be symmetric as well. In this paper we discuss two specific symmetries, both leading to a reduction of the number of variables by a factor of two, and reduction of the number of constraints by a factor of two as well. While identifying the variables that can be removed is often easy, identification of the appropriate constraints can be harder and non-trivial. Each of these two symmetries therefore reduces the size of the problem matrix, and therefore memory consumption by a factor of four. While current-generation large-scale optimization methods do attempt to identify existing symmetries before starting the actual solving process, they fail to (fully) find the symmetries that we will present. Therefore, in most cases, we do see an improvement in computation time as well.

\subsection{Point symmetry}
\label{sec:point_symmetry}

Point symmetry occurs when our focal-plane mask $M(\vect{k})$, dark zone $D$ and contrast limit $c(\vect{k})$ are point-symmetric, and if our pupil and Lyot stops are real. That is,
\begin{subequations}
\begin{align}
M(-\vect{k}) &= M(\vect{k})~&~~\forall~\vect{k}, \\
-\vect{k} &\in D~&~~\forall~\vect{k}\in D, \\
c(-\vect{k}) &= c(\vect{k})~&~~\forall~\vect{k}.
\end{align}
\end{subequations}
In this case, the optimization problem is invariant under complex conjugation of the apodizer $\Phi(\vect{x}) \to \Phi^*(\vect{x})$, where $(\cdot)^*$ denotes complex conjugation. Therefore, our solution must be real: \begin{equation}
\Phi(\vect{x}) = \Phi^*(\vect{x}) \in \mathbb{R}.
\end{equation}
We can therefore remove all imaginary component from the apodizer mask $\Phi$, removing half of all variables in our optimization problem.

We can now find the result of this symmetry on our constraints. As our electric field at the apodizer plane is now real, our electric field at the focal-plane mask will Hermitian. Due to the point-symmetric focal-plane mask, the electric field after the focal-plane mask will also be Hermitian. Therefore, the electric field at the Lyot-stop plane will be real. As the Lyot-stop mask is real, the electric field after the Lyot-stop mask is also real. Therefore, we know that the electric field of our coronagraphic image has to be Hermitian symmetric:
\begin{equation}
\Psi^\mathrm{coro, aber}_{\lambda,L,\vect{\alpha}}(\vect{k}) = (\Psi^\mathrm{coro, aber}_{\lambda,L,\vect{\alpha}}(-\vect{k}))^*.
\end{equation}
As our constraints are symmetric under complex conjugation, we only need to add focal-plane constraints for one side of the dark zone. The electric field on the other side will automatically satisfy the focal-plane constraint without explicit inclusion. For simplicity, we often remove either the right or top half of the dark zone. The choice between these two is determined by the symmetry in the next section.

Note that we have not set any conditions on the Lyot stop or telescope pupil (other than that they are real). This means that this symmetry is applicable to all APLC designs that we want to design.

\begin{figure}
\centering
\includegraphics[width=0.85\textwidth]{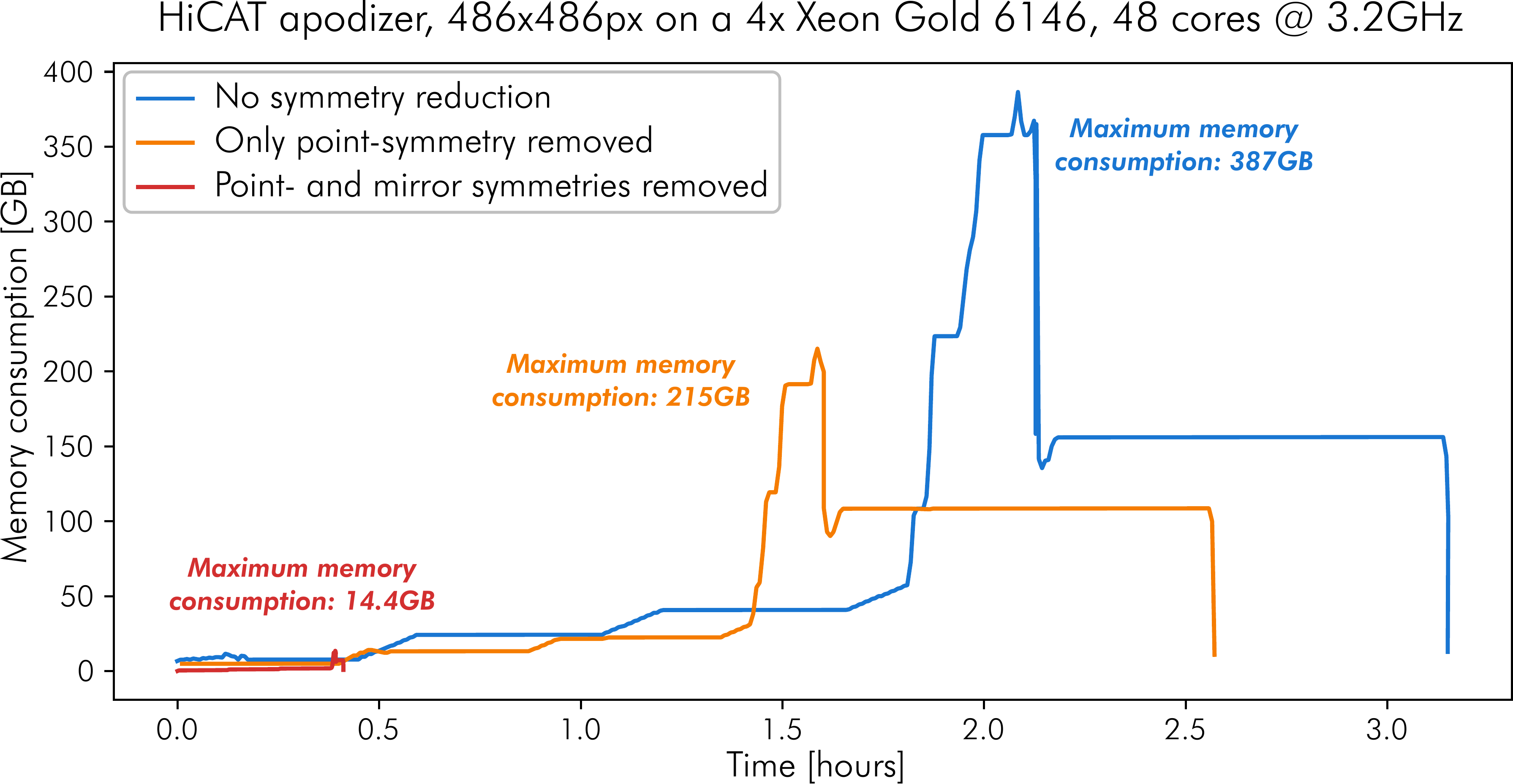}
\vspace*{2mm}
\caption{Memory consumption over time during optimization of an apodizer for HiCAT with different symmetry reductions disabled or enabled. The optimization was performed for the HiCAT pupil at 486x486px resolution, on a server with four Xeon Gold 6146's for a total of 48 cores running at 3.2GHz and having 512GB of memory. Reduction of the point symmetry reduces peak memory consumption by about two fold. Additionally removing all mirror symmetries gains another factor $\sim16\times$. All three optimizations yield solutions within the numerical accuracy of the optimizer.}
\label{fig:symmetry_memory_graph}
\end{figure}

In Figure~\ref{fig:symmetry_memory_graph} we show the memory consumption over time during the optimization of masks for the HiCAT testbed. We can see about a two-fold reduction in peak memory consumption when comparing the memory trace without any symmetry reduction and the trace with only point symmetry removed. While we would expect a four-fold reduction, the optimization with no symmetry reduction already removes the imaginary component of the apodizer mask as we are optimizing an APLC mask, which is required to be real. Therefore, we only see the two-fold reduction from the halving of the dark zone.

\begin{table}
\centering
\vspace{3mm}
\begin{tabular}{ccccc}
\toprule
\multicolumn{1}{c}{\begin{tabular}[c]{@{}c@{}}\textbf{Point symmetry}\\\textbf{removed?}\end{tabular}} & 
\multicolumn{1}{c}{\begin{tabular}[c]{@{}c@{}}\textbf{Mirror symmetries}\\\textbf{removed?}\end{tabular}} &
\multicolumn{1}{c}{\textbf{\# variables}} &
\multicolumn{1}{c}{\textbf{\# constraints}} &
\multicolumn{1}{c}{\begin{tabular}[c]{@{}c@{}}\textbf{Reduction in}\\\textbf{memory consumption}\end{tabular}} \\ \midrule
\xmark & \xmark & 141,716 & 31,872 & $0\%$ \\
\cmark & \xmark  & 141,716 & 15,936 & $\sim50\%$ \\
\cmark & \cmark & 35,429 & 3984 & $\sim97\%$ \\
\bottomrule
\end{tabular}
\vspace{2mm}
\caption{Number of variables and constraints for each of the optimizations shown in Figure~\ref{fig:symmetry_memory_graph}. Removing point symmetry from the optimization problem improves memory consumption by $2\times$, while additionally removing a two-fold mirror symmetry improves memory consumption by $32\times$.}
\label{tab:symmetry_memory_consumption}
\end{table}

\subsection{Mirror symmetry}
\label{sec:mirror_symmetry}

Mirror symmetries occur our pupil, focal-plane mask, Lyot-stop masks and dark zone exhibit mirror symmetries. Here we need to make the distinction between single and double mirror symmetric problems.

\subsubsection{Single mirror symmetry}

Without loss of generality, we can choose to have mirror symmetry along the x axis. Mirror symmetries along other lines can be handled by manually rotating the pupil and Lyot-stop masks beforehand. For our problem to exhibit mirror symmetry along x, we need to have a pupil and set of Lyot-stop masks that are mirror symmetric along x, and a focal-plane mask and dark zone that are mirror symmetric along y. Mathematically,
\begin{subequations}
\begin{align}
\Pi(-x, y) &= \Pi(x, y) &~~\forall~x,y\\
M(k, -l) &= M(k,l) &~~\forall~k,l\\
L_0(-x, y) &= L_0(x, y) &~~\forall~x,y\\
\{L_1(-x, y), L_2(-x, y), \ldots, L_N(-x, y)\} &= \{L_1(x, y), L_2(x, y), \ldots, L_N(x, y)\} &~~\forall~x,y,\\
(k, -l) &\in D&~~\forall~(k,l) \in D
\end{align}
\end{subequations}
and analogous for mirror symmetry in y. Note that the mirror symmetry for the set of Lyot stops is taken along the entire set. This means that individual Lyot stops are allowed to be asymmetric, as long as their counterpart is also an element of the set of Lyot stops. For example, while a Lyot stop that is slightly shifted towards positive x is not mirror-symmetric along x anymore, even if the nominal Lyot stop was mirror symmetric, as long as version that is shifted towards negative x is also an element of the set.

In this case, the optimization problem is invariant under Hermitian conjugation of the apodizer along the x-axis:
\begin{equation}
\Phi(x, y) = \Phi^*(-x, y)~\forall~x,y.
\end{equation}
This means that we can remove half of the variables for the apodizer mask.

Again, identifying which constraints to remove is non-trivial. For a mirror symmetry along x, the electric field at the focal-plane mask will be Hermitian along the y-axis:
\begin{equation}
\Psi_B(x, y) = \Psi^*_B(x, -y).
\label{eq:hermitian_symmetry_x_mirror}
\end{equation}
This is easiest to see when splitting the two-dimensional Fourier transform into two one-dimensional Fourier transforms, one along x, and one along y. For mirror symmetry along x, we first apply the one-dimensional Fourier transform along x, yielding a real function. Then, applying the one-dimensional Fourier transform along y, yields the one-dimensional Hermitian symmetry as in Eq.~\ref{eq:hermitian_symmetry_x_mirror}. As the focal-plane mask is real and mirror symmetric in y, this one-dimensional Hermitian symmetry is retained. Therefore, the electric field at the Lyot stop is again Hermitian symmetric along x.

Now for the Lyot stop we have two options:
\begin{enumerate}
\item The Lyot stop itself is mirror symmetric along x. In this case mirror symmetry is retained, and the field in the final focal plane will again exhibit one-dimensional Hermitian symmetry along y. Therefore, we can choose to omit half of the focal-plane constraints, either the top or the bottom half. Note that if point symmetry also applies, as is often the case, we are free to choose any quarter of the dark zone. This is because we are free to choose either horizontal half for the point symmetry reduction. If our problem exhibits mirror symmetry along y instead of x, then we are free to choose the top half for the point symmetry, and the left half for the mirror symmetry. In both cases, we are left with a quarter of the dark zone, which may seem counter intuitive at first.
\item The Lyot stop itself is not mirror symmetric along x, but the set of Lyot stops contains its symmetric counterpart. In this case, the electric field after the Lyot stop will not be symmetric anymore, along with the electric field at the final focal plane. However, between the asymmetric Lyot stop and its counterpart, we can see that this amounts to a reflection along x. Therefore, in the final focal plane, the transformation from using the asymmetric Lyot stop to its counterpart, is equal to complex conjugation of the field and mirror reflection along y. Therefore, as our focal-plane constraints are symmetric in real and imaginary part, we can omit one of the Lyot stops completely.
\end{enumerate}

Therefore, in both cases we remove half of the focal-plane constraints. Along with the removal of variables, a single mirror symmetry yields a four-fold reduction in memory consumption.

\subsubsection{Two-fold mirror symmetry}

For problems that exhibit two-fold mirror symmetry, the situation becomes slightly different. The conditions for this case are
\begin{subequations}
\begin{align}
\Pi(-x, y) &= \Pi(x, -y)=\Pi(x, y) &~~\forall~x,y\\
M(-k, l) &= M(k,-l)=M(k, l) &~~\forall~k,l\\
L_0(-x, y) &= L_0(x, -y) = L_0(x, y) &~~\forall~x,y\\
\nonumber\{L_1(-x, y), L_2(-x, y), \ldots, L_N(-x, y)\} &= \{L_1(x, -y), L_2(x, -y), \ldots, L_N(x, -y)\}\\
&= \{L_1(x, y), L_2(x, y), \ldots, L_N(x, y)\} &~~\forall~x,y,\\
(-k, l) &\in D&~~\forall~(k,l) \in D\\
(k, -l) &\in D&~~\forall~(k,l) \in D.
\end{align}
\end{subequations}
Again, the Lyot-stop masks have to be symmetric as a set and not necessarily individually. Under these conditions, the problem is invariant under either Hermitian conjugation of the apodizer along x and y, yielding
\begin{equation}
\Phi(x, y) = \Phi^*(-x, -y) = \Phi^*(x, -y) = \Phi(-x, -y)~~\forall~x,y.
\end{equation}
This yields a removal of \sfrac34 of all variables in the apodizer mask.

Again, identifying which constraints can be omitted is non-trivial. The electric field at the focal-plane mask will be two-fold one-dimensional Hermitian. Again, this is easier to see after splitting the two-dimensional Fourier transforms into two one-dimensional Fourier transforms. After the first one-dimensional Fourier transform along x, the field will be real, and the second one-dimensional Fourier transform will make it Hermitian along y. Alternatively, we could perform the one-dimensional Fourier transform along y first, and then along x. This yields a one-dimensional Hermitian function along x. Therefore, the final function must be two-fold one-dimensional Hermitian. As the focal-plane mask is real and two-fold mirror symmetric, it does not affect this symmetry. Therefore, the electric field at the Lyot stop is again two-fold one-dimensional Hermitian.

We split up the Lyot stop into three cases:
\begin{enumerate}
\item The Lyot stop itself is two-fold mirror symmetric. In this case, the final focal plane will be two-fold one-dimensional Hermitian as well, and we can omit \sfrac34 of the dark zone. Note that if point symmetry also applies, the field will be purely real as well, and we only need to add the constraints for the real part for a quarter of the dark zone.
\item The Lyot stop itself is only mirror symmetric along one axis, but its counterpart for the other mirror symmetry is an element of the set of Lyot stops. Let's assume that Lyot stop itself is mirror symmetric in y and that its counterpart is its mirror reflection along x. In this case, analogous to the case for single mirror symmetry, we can retrieve the field for the other Lyot stop, we need to take the Hermitian conjugate along the y axis. Therefore, As our focal-plane constraints are mirror symmetric, we can safely omit one of the Lyot stops completely, and furthermore use only half of the dark zone. Note that we still need to constrain both real and imaginary components. Note that if point symmetry also applies, we are free to only constrain a quarter of the dark zone.
\item The Lyot stop itself is not mirror symmetric over either of the two axes, but its three counterparts are all elements of the set of Lyot stops. In this case, analogous to the case above, we can remove the three Lyot stops from the optimization problem. We however do need to constrain both real and imaginary parts for the full dark zone. Note that if point symmetry also applies, we can still remove any half of the dark zone here.
\end{enumerate}

The red line in Figure~\ref{fig:symmetry_memory_graph} depicts the memory consumption over time for a two-fold mirror symmetric problem with the HiCAT aperture. We can see that we gain another $\sim4\times4=\sim16\times$ reduction in peak memory consumption compared to the purely point-symmetric case. In total, we achieve an improvement of $\sim4\times4\times2=\sim32\times$ reduction in memory consumption.

\section{Progressive refinement}
\label{sec:progressive_refinement}

\subsection{General idea}

It is well known that solution of APLC apodizers are binary, consisting of patches of zero and one transmission\cite{zimmerman2016shaped}. When performing the same optimization at different resolutions, we notice that the patches do not move or change shape, and that just the edges get more defined and sharper as the resolution increases. Additionally, low resolution optimizations that resolve individual patches are useful for probing the parameter space: even though the mask itself may not have a sufficiently high resolution to be manufactured, it still gives realistic values for coronagraphic throughput. Therefore, these low resolution optimizations are often used to perform parameter studies for determining the best hyperparameters. Think here for example of the focal-plane mask radius, the inner and outer diameter of the Lyot stop and thickness of the spiders in the Lyot stop. For probing the full parameter space we would like to perform thousands of optimizations, which is only possible at low resolutions.

Here we aim to use low-resolution optimization for another purpose. As each low-resolution mask already contains roughly the correct size and shape for each of the patches in the apodizer, we would like to use this information as a prior for higher resolution optimizations. The middle of these patches are unlikely to change across different resolutions. Therefore we can imagine fixing these pixels in a high resolution optimization and only optimize pixels at the edge of the patches. This type of technique is well known to the computer graphics community as ``progressive refinement''. Progressive refinement starts off with a low resolution estimate of an image, providing the user with a coarse estimation of the final image. This is useful for quick checks and can allow early cancellation of the computation if required. This coarse image is then gradually improved upon as more data becomes available or is computed. 

As the number of pixels on the edge is much smaller than the number of pixels in the whole apodizer, we can in this way remove most of the variables from high resolution optimizations, while only having to perform a comparatively quick low-resolution optimization beforehand. Furthermore, the number of pixels on the edges of patches grows linearly with the resolution of the apodizer rather than quadratically. Therefore the improvement in memory consumption will become better then larger the apodizer is that we want to optimize.

\subsection{Example}

\begin{figure}
\centering
\includegraphics[width=0.85\textwidth]{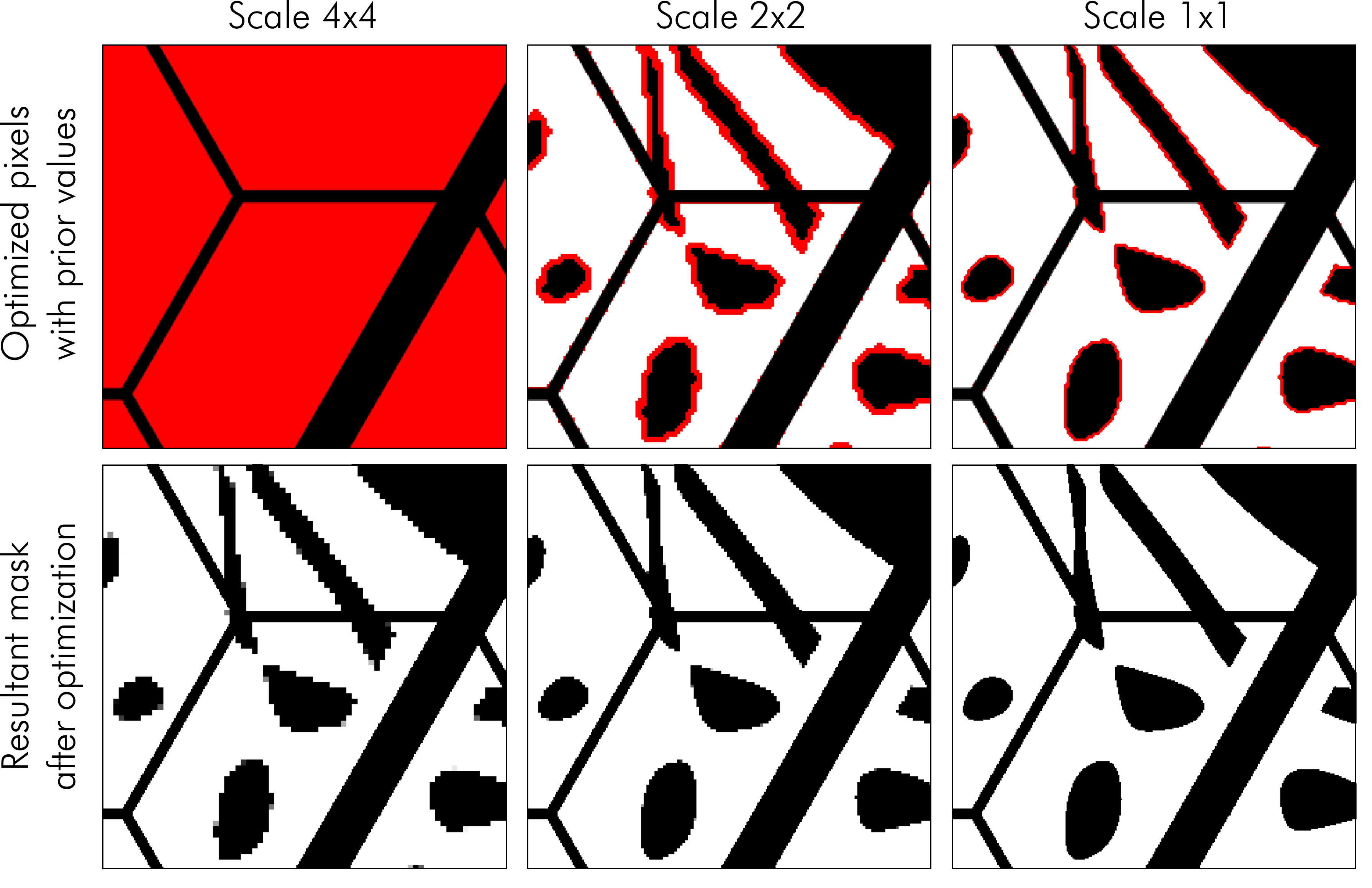}
\vspace*{3mm}
\caption{Example of the optimized pixels and the resulting masks at different resolutions with the progressive refinement algorithm. For the top row: red indicates that a pixel is optimized at that resolution, and white/black indicates the value for pixels that are not optimized.}
\label{fig:progressive_masks}
\end{figure}

In Figure~\ref{fig:progressive_masks} we show this technique in action. To show the details in the optimized apodizer, we only show a zoomed-in version of the full mask. We start off at a scale of $486\times486\mathrm{px}$. As we have not performed any optimization, every pixel needs to be optimized at this resolution. Pixels that are optimized are shown in red, while pixels that are not optimized are shown either in black or white depending on the fixed value at the current resolution. At this resolution we retrieve the mask shown on the bottom left.

Upscaling this mask to twice its original resolution ($972\times972\mathrm{px}$), we only need to optimize pixels that lay on the edge of patches. We use an edge detection algorithm to automatically identify the edges, and use binary dilation to thicken the edges to 3-5 pixels wide. This gives enough room for edges to move slightly, while still not degrading performance for no reason. Additionally we include pixels that were not sufficiently black or white in the low-resolution optimization. These pixels often indicate unresolved structure in the apodizer that might become resolved at higher resolutions. This is often the case at the edges of the Lyot stop projected onto the apodizer, in cases where we include Lyot robustness. The result of this optimization with prior is shown in the bottom middle. Finally, this process can be repeated a third time, upscaling to a resolution of $1944\times1944\mathrm{px}$, with the results in the right column. 

\subsection{Analysis}

\begin{figure}
\centering
\includegraphics[width=0.85\textwidth]{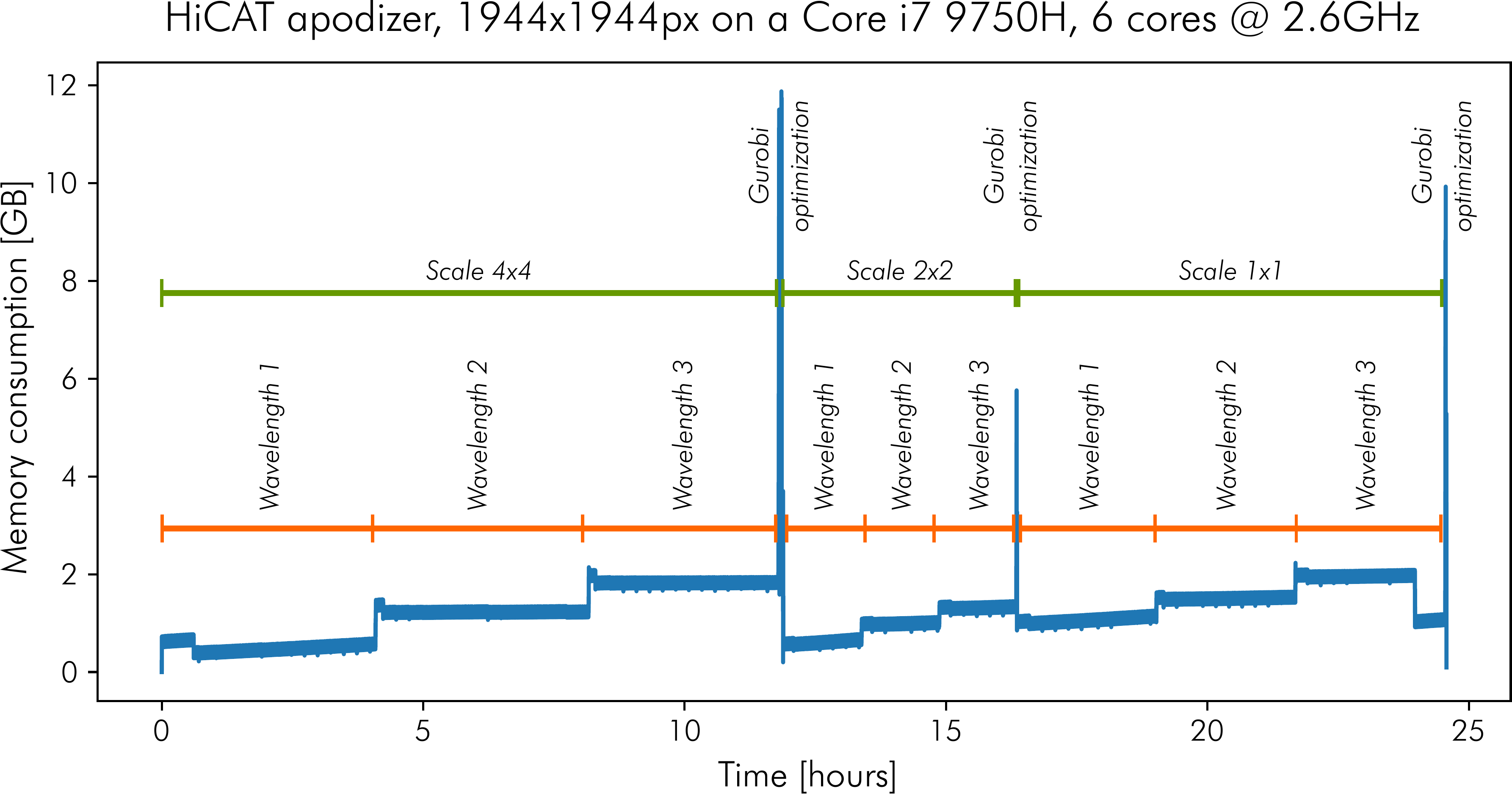}
\vspace*{2mm}
\caption{Memory consumption over time during the optimization of Fig.~\ref{fig:progressive_masks}. The different phases during the optimization process are annotated. In total, four numerical optimization problems were solved. The first two are around $11.5\mathrm{hours}$, which includes an optimization for peak throughput convergence. The two other occur at $\sim16.5\mathrm{hours}$ and $24.5\mathrm{hours}$.}
\label{fig:progressive_memory_graph}
\end{figure}

Figure~\ref{fig:progressive_memory_graph} shows the memory consumption over time for the full optimization, starting at \sfrac14-th scale to full resolution. The graph is annotated to show different stages of the optimization process, starting with calculation of the problem matrix for each of the three wavelengths at \sfrac14-th scale, then optimizing the apodizer at that scale, and then doing the same for \sfrac12 scale and full resolution. The optimization ran on a laptop equipped with a Core i7 9750H with six cores running at 2.6GHz.

Calculation of the problem matrix is performed using propagations at the full, final resolution, using superpixels of $2^n\times2^n\mathrm{px}$ depending on the current scale. This is a very conservative approach, which makes sure that superpixels that intersect the edge of the telescope pupil are handled in the same way throughout the optimization process. This conservative approach is the reason for the large amount of time spent on the problem matrix calculation compared to Figure~\ref{fig:symmetry_memory_graph}, where the two-fold symmetric (red) line depicts an optimization of the same complexity, albeit on different, faster hardware. In the future, we will experiment with performing this calculation at the lower resolution.

Table~\ref{tab:progressive_memory_consumption} shows the number of variables and constraints at each of the stages of the progressive refinement algorithm. We can see that the progressive refinement algorithm is able to remove $\sim90\%$ of all variables at the $2\times2$ scale, and $\sim95\%$ at the final resolution. Note that from $2\times2$ to $1\times1$ the number of variables in the progressive algorithm only grew by a factor of $\sim2$, compared to $\sim4\times$ for the non-progressive algorithm. This illustrates that the progressive algorithm scales linearly with resolution rather than quadratic. The final optimization stage achieved a peak memory consumption of $\sim10\mathrm{GB}$ which would have required $\sim200\mathrm{GB}$ without progressive refinement. This would have necessitated the use of a high-memory server instance. Additionally, these memory figures already include the gains made in Section~\ref{sec:removing_symmetries}. 

\begin{table}[t]
\centering
\vspace{3mm}
\begin{tabular}{ccccc}
\toprule
\multicolumn{1}{c}{\textbf{Scale}} & \multicolumn{1}{c}{\begin{tabular}[c]{@{}c@{}}\textbf{\# variables}\\\textbf{(progressive)}\end{tabular}} & \multicolumn{1}{c}{\begin{tabular}[c]{@{}c@{}}\textbf{\# variables}\\\textbf{(non-progressive)}\end{tabular}} & \multicolumn{1}{c}{\textbf{\# constraints}} & \multicolumn{1}{c}{\begin{tabular}[c]{@{}c@{}}\textbf{Reduction in}\\\textbf{memory consumption}\end{tabular}} \\ \midrule
$4\times4$ & 35,429 & 35,429 & 3,984 & $0\%$ \\
$2\times2$ & 14,213 & 141,716 & 3,984 & $\sim90\%$ \\
$1\times1$ & 25,691 & 566,864 & 3,984 & $\sim95\%$ \\ \bottomrule
\end{tabular}
\vspace{2mm}
\caption{Number of variables and constraints at each scale during the progressive refinement algorithm ran for Figure~\ref{fig:progressive_memory_graph}. At the lowest resolution $4\times4$ no improvement is memory consumption is observed as all pixels are optimized at the initial scale. The number of variables is listed for both progressive and non-progressive optimization. For the $2\times2$ scale we are able to remove $\sim90\%$ of all variables, and $\sim95\%$ at the final full resolution.}
\label{tab:progressive_memory_consumption}
\end{table}

In some cases we have seen that the optimizer is unable to find a feasible solution. This means that we were too aggressive with the removal of variables, and that with the available pixels the optimizer was unable to achieve the required contrast limit. In these cases we can either restart the progressive refinement and allow more pixels to be optimized along the edges, broadening the band of pixels that are allowed to be optimized. If this also fails, which we have seen in some rare cases, the optimization must be restarted with the current scale as the initial scale, ie. allowing all pixels to partake in the optimization.

\section{Case studies}
\label{sec:case_studies}

We developed a code that implements the reductions outlined in the previous sections. The code automatically determines which symmetries are applicable to the current problem, and removes the appropriate constraints and variables from the optimization problem. Simultaneously, it can perform progressive refinement on the found solutions, iteratively upscaling lower-resolution solutions. The simulation of all propagations is done with HCIPy,  an open-source object-oriented framework written in Python for performing end-to-end simulations of high-contrast imaging instruments\cite{por2018hcipy}. The Lyot coronagraph in HCIPy uses the semi-analytical method\cite{soummer2007fast}. For efficiently reasons, the code only propagates light to the part of the dark zone that is used for the constraints. The numerical optimization itself is performed by Gurobi\cite{gurobi}.

The code contains a survey mode that simplifies running large parameter studies. This includes logging of all relevant information, and producing a paper trail for the whole optimization so that the exact optimization can be repeated and/or checked in the future. Additionally, the code automatically produces a PDF-file of relevant plots so that a design can be evaluated at first glace immediately after optimization.

The code was used to produce designs for HiCAT\cite{soummer2018hicat5} and LUVOIR-A\cite{pueyo2017luvoir}. This section will highlight some of the optimized designs.

\subsection{HiCAT}

\begin{figure}
\centering
\includegraphics[width=0.95\textwidth]{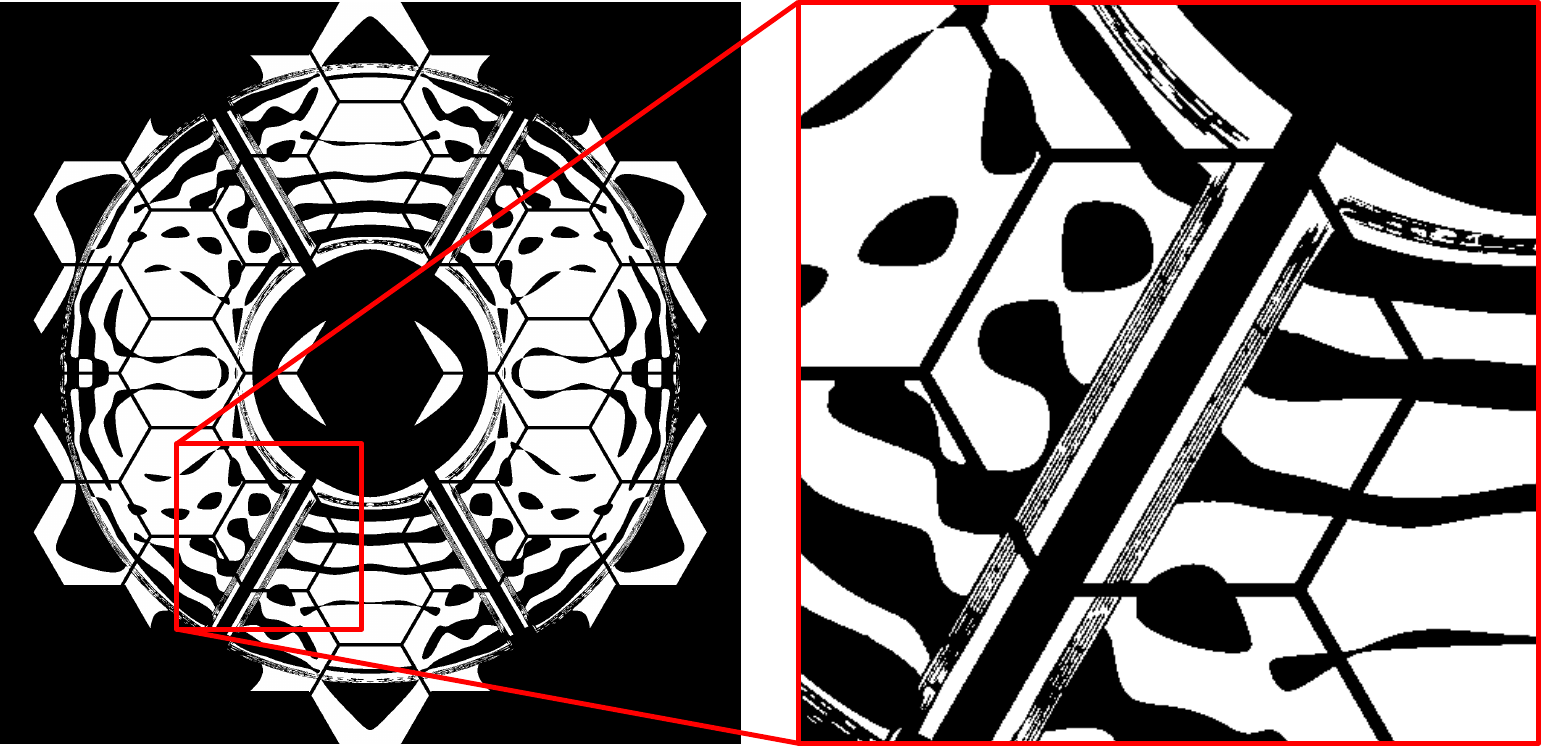}
\vspace{2mm}
\caption{The optimized mask for the HiCAT testbed. These masks were optimized at the manufacturing resolution and contain a $\pm0.3\%$ robustness to Lyot stop translations. This translation range can clearly be seen as the optimizer adds high frequency features in the apodizer design to accommodate these different Lyot stop translations.}
\label{fig:hicat_mask}
\end{figure}

The HiCAT pupil is two-fold mirror symmetric and therefore takes the greatest advantage of the improvements shown in this paper. Without the improvements in this paper, this optimization would have required an estimated $63\mathrm{TB}$ of memory. In Figure~\ref{fig:hicat_mask} we show the optimized design, optimized at the manufactured resolution of $1944\times1944\mathrm{px}$. Lyot robustness is included in the optimizations with a set of nine Lyot stops in a three-by-three grid centered around the nominal position.

\begin{figure}
\centering
\includegraphics[width=0.63\textwidth]{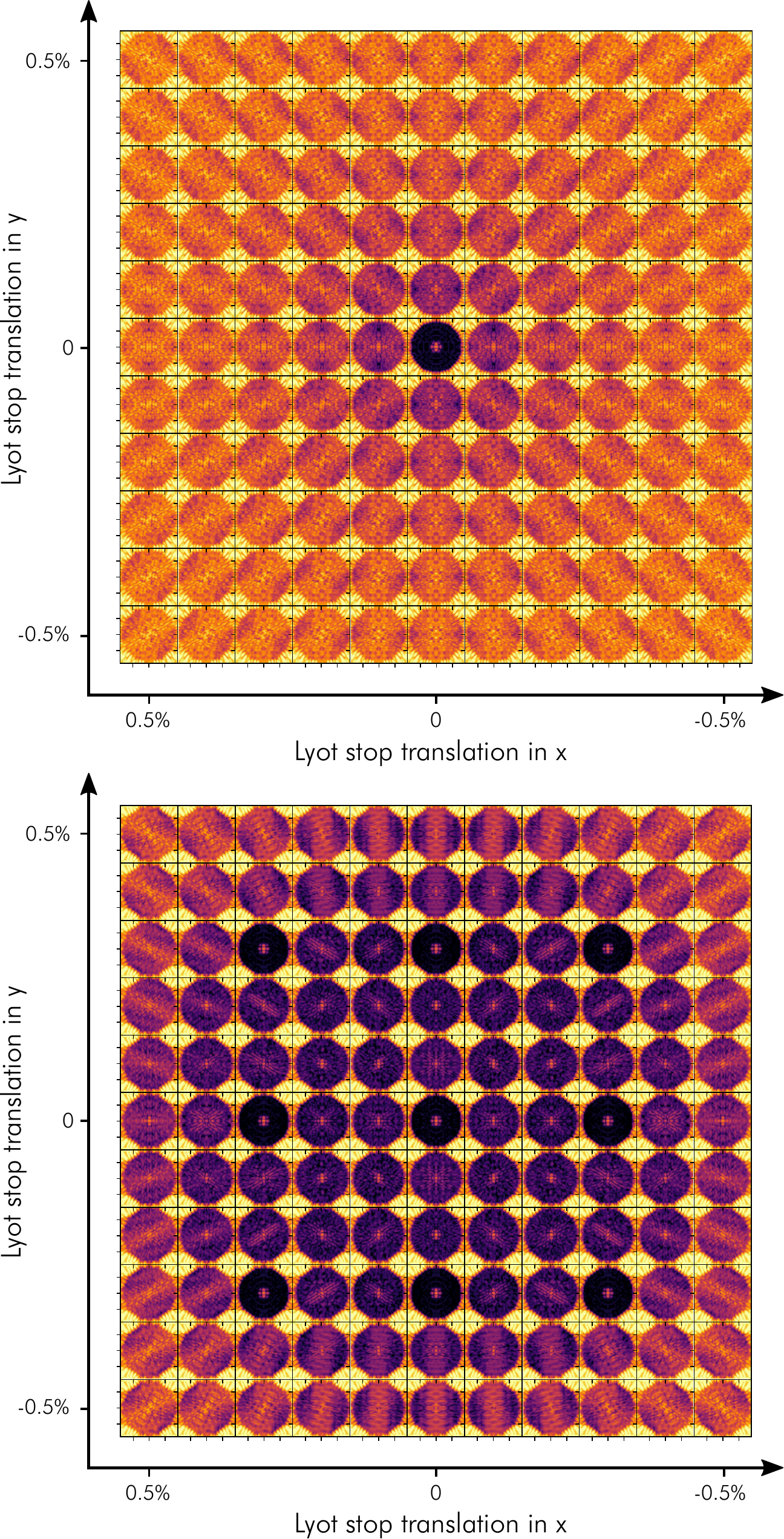}
\vspace{2mm}
\caption{Two analysis plots for two different HiCAT apodizer designs, one without and one with added robustness against Lyot stop translations. The colorbar for all images are the same, stretching in logarithmic scale from $10^{-4}$ to $10^{-9}$.}
\label{fig:hicat_lyot_sensitivity}
\end{figure}

The improved speed of the optimizer allows for better comparison of different strategies of adding robustness to Lyot stop translations. Figure~\ref{fig:hicat_lyot_sensitivity} shows the Lyot sensitivity for a mask with and without built-in Lyot robustness. These figures show the coronagraphic image for different translations of the Lyot stop mask. Clearly, the design that is made robust against Lyot stop translations is more capable of handling these small perturbations. The optimization is performed with a set of nine Lyot stops, which are still clearly visible in the grid of Lyot stop translations as overall darker images.

\subsection{LUVOIR-A}

\begin{figure}
\centering
\includegraphics[width=0.8\textwidth]{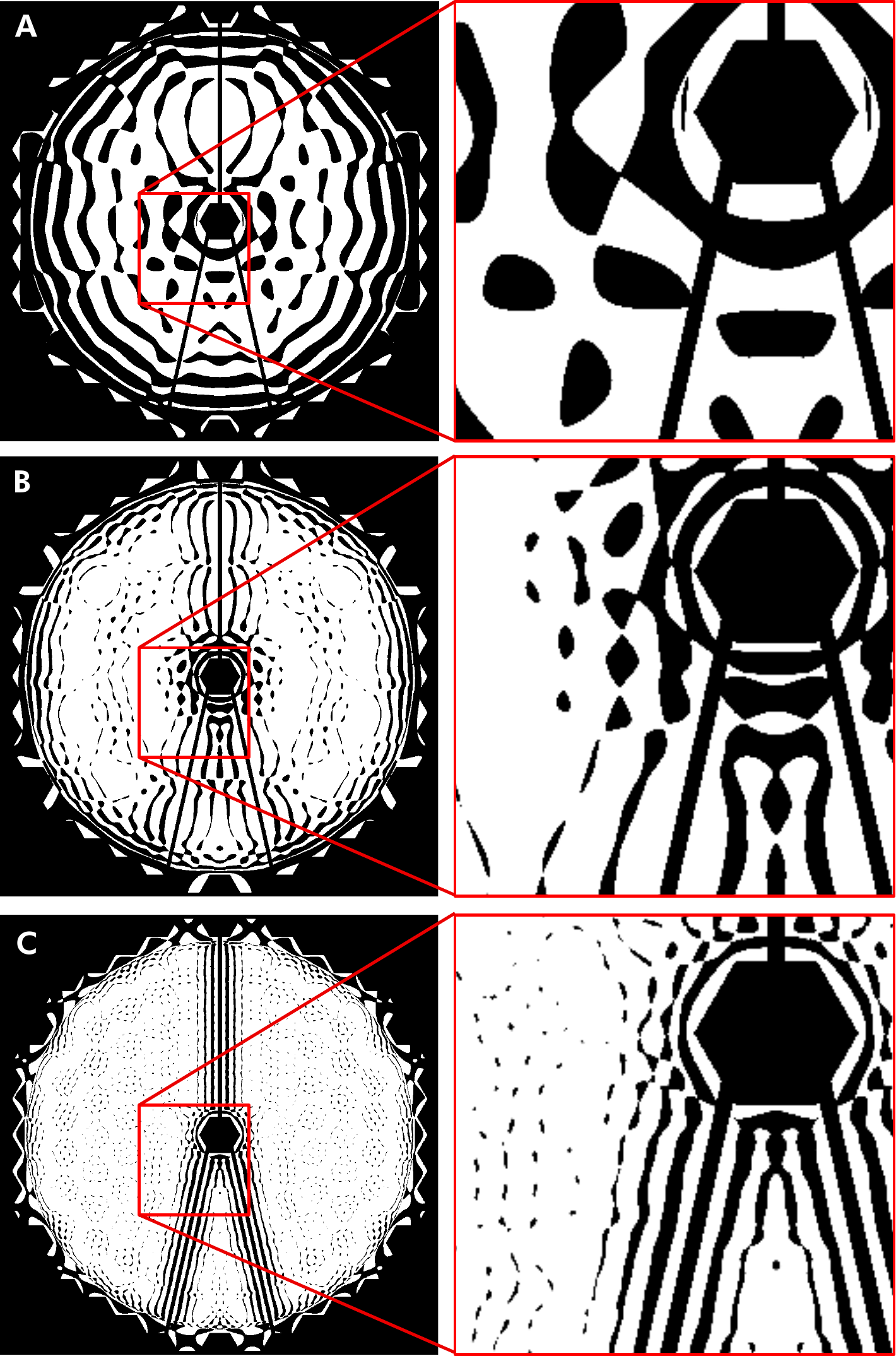}
\vspace{2mm}
\caption{The three masks optimized for the LUVOIR-A telescope pupil, with a small, medium and large inner working angle to provide the best mask for different scenarios.}
\label{fig:luvoir_masks}
\end{figure}

The LUVOIR-A pupil is only mirror symmetric along x and our optimizations are therefore performed at a reduced resolution of  $1024\times1024\mathrm{px}$. The current concept for the APLC in LUVOIR-A calls for three masks with differently sized focal-plane masks and outer working angles. These masks are shown in Figure~\ref{fig:luvoir_masks}. While the apodizer paired with the smallest focal-plane mask can be used to look for planets very close in, it does not have the best coronagraphic throughput when this capability is not required. Having a larger focal-plane mask allows us to improve coronagraphic throughput to aid in efficient characterization of exoplanets in wider orbits.

\section{Conclusions}
\label{sec:conclusions}

In this paper we described two ways to reduce the memory consumption when optimizing APLC apodizer masks. The first exploited symmetries in the underlying problem, specifically point symmetry and the two mirror symmetries. The second way exploited the binary structure of the masks to progressively upscale low-resolutions at minimal cost using progressive refinement.

We showed two examples of these reductions in action, optimizing apodizer masks for the HiCAT testbed at the native manufacturing resolution, and for the LUVOIR-A telescope. The HiCAT apodizer mask would have required an estimated $63\mathrm{TB}$ of memory without the improvements in this paper. The LUVOIR-A apodizer with the largest outer working angle would have required an estimated $22\mathrm{TB}$ of memory. This shows that much larger and more intricate masks can be optimized with these improvements, allowing for better exploration of the parameter space and easier optimization at native manufacturing resolutions.

\acknowledgments
EHP acknowledges funding by The Netherlands Organisation for Scientific Research (NWO) and the S\~{a}o Paulo Research Foundation (FAPESP). EHP also thanks STScI for their hospitality during his visit in April 2019 when most of the work in this paper was performed. This work was also supported in part by the Segmented-aperture Coronagraph Design and Analysis funded by ExEP, under JPL subcontract No.1539872, and in part by the National Aeronautics and Space Administration under Grant 80NSSC19K0120 issued through the Strategic Astrophysics Technology/Technology Demonstration for Exoplanet Missions Program (SAT-TDEM; PI: R. Soummer).

\bibliography{report}
\bibliographystyle{spiebib}

\end{document}